\documentclass[twocolumn,english,aps,prb,showpacs,superscriptaddress]{revtex4}
\usepackage[T1]{fontenc}
\usepackage[latin9]{inputenc}
\usepackage{color}
\usepackage{amsmath}
\usepackage{amssymb}
\usepackage{stackrel}
\usepackage{graphicx}
\usepackage{ulem}
\usepackage{float}
\usepackage{lipsum}
\usepackage{babel}
\usepackage{epstopdf}

\makeatother

\usepackage{babel}
\begin{document}
\title{Exponential size scaling of the Liouvillian gap in boundary-dissipated
systems with Anderson localization}
\author{Bozhen Zhou}
\thanks{These authors contributed equally to this work.}
\affiliation{Beijing National Laboratory for Condensed Matter Physics, Institute
of Physics, Chinese Academy of Sciences, Beijing 100190, China}
\affiliation{School of Physical Sciences, University of Chinese Academy of Sciences,
Beijing 100049, China }
\author{Xueliang Wang}
\thanks{These authors contributed equally to this work.}
\affiliation{Beijing National Laboratory for Condensed Matter Physics, Institute
of Physics, Chinese Academy of Sciences, Beijing 100190, China}
\affiliation{School of Physical Sciences, University of Chinese Academy of Sciences,
Beijing 100049, China }
\author{Shu Chen}
\thanks{Corresponding author: schen@iphy.ac.cn }
\affiliation{Beijing National Laboratory for Condensed Matter Physics, Institute
of Physics, Chinese Academy of Sciences, Beijing 100190, China}
\affiliation{School of Physical Sciences, University of Chinese Academy of Sciences,
Beijing 100049, China }
\affiliation{Yangtze River Delta Physics Research Center, Liyang, Jiangsu 213300,
China }
\date{\today}
\begin{abstract}

We carry out a systematical study of the size scaling of Liouvillian gap in boundary-dissipated one-dimensional quasiperiodic and disorder systems.
By treating the boundary-dissipation operators as a perturbation, we derive an analytical expression of the Liouvillian gap, which indicates clearly the Liouvillian gap being proportional to the minimum of boundary densities of eigenstates of the underlying Hamiltonian, and thus give a theoretical explanation why the Liouvillian gap has different size scaling relation in the extended and localized phase.
While the Liouvillian gap displays a power-law size scaling $\Delta_{g}\propto L^{- 3}$  in the extended phase, our analytical result unveils that the Liouvillian gap
fulfills an exponential scaling relation $\Delta_{g}\propto e^{- \kappa L}$
in the localized phase, where $\kappa$ takes the largest Lyapunov exponent of localized eigenstates of the underlying Hamiltonian. By scrutinizing the extended Aubry-Andr\'{e}-Harper
model, we numerically confirm that the Liouvillian gap fulfills the exponential scaling relation and the fitting exponent $\kappa$  coincides pretty well
with the analytical result of Lyapunov exponent.
The exponential scaling relation is further verified numerically in other
one-dimensional quasiperiodic and random disorder models. We also
study the relaxation dynamics and show the inverse of Liouvillian gap
giving a reasonable timescale of asymptotic convergence to the steady
state.
\end{abstract}
\maketitle

\section{Introduction}


In the past years, advances in manipulating dissipation and quantum
coherence in laboratory have led to a renewed interest in the study
of open quantum systems with intriguing dissipative dynamics \citep{Weimer2021RMP,Diehl2008,Cirac2009,CaiZ,Prosen2012,Zhong2019PRL,LiuCHPRR,Poletti,Prosen2008NJP,Ueda}.
Understanding dynamical processes evolving to steady states in open quantum
systems driven by boundary dissipations is a central problem of out-of-equilibrium statistical physics attracted intensive theoretical studies \cite{Prosen2008PRL,Prosen2014PRL,Znidaric2015PRE,GuoC2018,Yoo,Popkov,Lacerda,Shibata2020PTEP,Popkov2013,Vicari,Znidaric2021,Znidaric2017,Schulz,Carollo,Briegel2013PRE,Znidaric2010JSM,Popkov2016PRA,ZhouZW,Monthus,Schaller2021arXiv} 
Within the Markovian approximation, 
the density matrix of the system evolves according to the Lindblad
master equation with the Liouvillian gap $\Delta_{g}$ defined as the smallest modulus
of the real part of nonzero eigenvalues of the Liouvillian superoperator. Usually, the inverse of the Liouvillian gap gives an estimation on the timescale of the relaxation time \cite{CaiZ,Prosen2008NJP,Znidaric2015PRE}. Although discrepancy between the inverse of Liouvillian gap and the relaxation time is found in some recent works \cite{Ueda,Znidaric2015PRE,Mori,Mori2021PRR,Bensa2022PRR}, the Liouvillian gap is still an important quantity characterizing the asymptotic convergence to the steady state \cite{Mori,Schaller2021arXiv,DengDL}.
Numerical results have demonstrated that the Liouvillian gap scales
with the system length $L$ in terms of $L^{-z}$ for various boundary-dissipated
systems \cite{Znidaric2015PRE,Shibata2020PTEP,Mori,Prosen2008PRL},
where $z\in[1,2)$ for chaotic systems and $z=3$ for integrable systems.

While most previous studies focus on the homogeneous systems, less
is known for the relaxation dynamics in disorder systems with boundary
dissipation. As localization has been recognized as important physical
implication of interference of waves in dissipative media, recently
there is growing interesting in the disorder effect on non-Hermitian
physics \cite{Hatano,ZengQB,JiangH,Longhi,LiuYX,Hughes,Ryu,ZhangDW,XuY}
and open quantum systems \cite{Denisov,Luitz,Can}, as well as the dynamical effect of Anderson localization induced by the Markovian noise \cite{Lorenzo,Lezama}. 
In Ref.\cite{Prosen2008NJP}, Prosen has provided
numerical evidence that the Liouvillian gap of the boundary-dissipated
disordered XY chain is exponentially small, i.e., $\Delta_{g}\propto e^{-L/\ell}$
with $\ell$ being the localization length of normal master mode.
Although the numerical result in Ref.\cite{Prosen2008NJP} suggests
that the Liouvillian gap should fulfill an exponential scaling relation
with the system length, a theoretical analysis and systematic study
of the Liouvillian gap for disorder systems with boundary dissipations
are still lacking. For a 1D disordered system, the localization length
of a localized eigenstate is usually energy dependent, and thus the
localization length of normal master mode is expected to be mode dependent,
so the meaning of $\ell$ is somewhat ambiguous.
Natural questions arising here are how to understand the role of normal
master modes in the formation of the Liouvillian gap and the connection
of Liouvillian gap to the localization lengths of eigenstates of the
underlying disordered chain?


To understand how the Liouvillian gap is affected
by the disorder, we first carry out a perturbative calculation by
treating the boundary-dissipation operators as a perturbation and
give an analytical derivation of the Liouvillian gap on the basis
of perturbation theory. Our analytical result indicates that the size
of Liouvillian gap is proportional to the minimum of boundary densities of eigenstates of the underlying Hamiltonian, and thus
the Liouvillian gap displays an exponential size scaling when the
underlying system possesses localized eigenstates. To get an intuitive
understanding from concrete examples, we then study the scaling relation
of Liouvillian gap numerically for various one-dimensional quasiperiodic
and disorder systems with boundary dissipations described by the Lindblad
master equation. The first example we consider is the extended Aubry-Andr\'{e}-Harper
(AAH) model with boundary dissipations. One of the reason for choosing
the extended AAH model is that it exhibits rich phase diagram with
extended (or delocalized), critical and localized phases depending on the quasiperiodical
modulation parameters \cite{Hatsugai,Takada,HanJH,WangYC}, and the
other reason is that the Lyapunov exponent (inverse of the localization
length) of the localized eigenstate of the model has an analytical
expression which is very helpful for checking our numerical fitting
results. Our numerical results illustrate that Liouvillian gap $\Delta_{g}$
displays different features in the underlying distinct phase regions.
While $\Delta_{g}\propto L^{-3}$ in the extended phase, the Liouvillian
gap scales with $L$ in an exponential way $e^{-aL}$ in the localized
phase, where $a$ is identified to be identical to the Lyapunov exponent
$\kappa$ of the localized state. 
To confirm the validity of the exponential scaling relation, we further
study a quasiperiodical model with mobility edge and the 1D Anderson
lattice, in which the localization length of a
localized eigenstate is energy dependent. Our numerical results show
that the Liouvillian gap displays similar exponential scaling relation
$e^{-aL}$ with $a$ determined by the Lyapunov exponent of states in the band edges.

The rest of paper is organized as follows. In
Sec. II A, we introduce the formalism for the calculation of Liouvillian
gap and present the analytical derivation of Liouvillian gap in the
scheme of perturbation theory. In Sec.II B, we first study the scaling relation of Liouvillian
gap in the boundary-dissipated
extended AAH model, and then extend our study to the boundary-dissipated
quasiperiodic model with mobility edge and the 1D Anderson model.
In Sec.II C, we discuss the relaxation time by numerically studying
the dynamical evolution of average occupation number. A summary is
given in the last section.

\section{Formalism, models and results}


\subsection{Formalism and perturbative calculation of Liouvillian gap}

We consider open systems with the dissipative dynamics of density
matrix $\rho(t)$ governed by the Lindblad master equation \cite{Lindblad,GKS}:
\begin{align}
\frac{d\rho}{dt}=\mathcal{L}\left[\rho\right]=-i\left[H,\rho\right]+\sum_{\mu}\left(2L_{\mu}\rho L_{\mu}^{\dagger}-\left\{ L_{\mu}^{\dagger}L_{\mu},\rho\right\} \right),\label{eq:Lind_MAT}
\end{align}
where $H$ is the Hamiltonian governing the unitary part of dynamics
of the system and $L_{\mu}$ are the Lindblad operators describing
the dissipative process 
with the index $\mu$ denoting the dissipation channels. Particularly,
we consider the boundary-dissipated systems with the Lindblad operators
acting only on the first and the last site of the lattice and taking
the form of
\begin{equation}
L_{1}=\sqrt{\gamma_{1}}c_{1},\,\,\,\,\,\,\,\,\,\,\,\,L_{L}=\sqrt{\gamma_{L}}c_{L},\label{eq:LindO}
\end{equation}
where $c_{j}$ is the fermion annihilation operator acting on the
site $j$ and $\gamma_{1}$ ($\gamma_{L}$) denotes the boundary dissipation
strength. In this work, we shall consider 1D quasiperiodic and disorder
fermion systems with quasiperiodic or random on-site potentials described
by the Hamiltonian

\begin{equation}
H=\sum_{i=1}^{L-1}J_{i}(c_{i}^{\dagger}c_{i+1}+c_{i+1}^{\dagger}c_{i})+\sum_{i=1}^{L}V_{i}c_{i}^{\dagger}c_{i},\label{eq:H_fermi-1}
\end{equation}
where $J_{i}$ represents the hopping amplitude between the $i$-th
and $(i+1)$-th sites and $V_{i}$ denotes the chemical potential
on the $i$-th site. Since the Hamiltonian is quadratic in fermionic
operators, Eq. (\ref{eq:Lind_MAT}) with linear dissipations also
takes a quadratic form. For a quadratic open fermionic model with
$L$ sites, solving for the Liouvillian gap of the
quantum Lindblad equation can be reduced to the diagonalization of
a $4L\times4L$ antisymmetric matrix \citep{Prosen2008NJP} or $L\times L$
non-Hermitian matrix \citep{Poletti,Zhong2019PRL}.

In Ref.\cite{Zhong2019PRL}, it is shown that the Liouvillian gap
can be obtained by
\begin{equation}
\Delta_{g}=\min[2\text{Re}(-\beta_{n})],
\end{equation}
where $\beta_{n}$ is the eigenvalue of damping matrix given by \citep{Zhong2019PRL}
\begin{equation}
X=ih^{T}-\left(M_{1}+M_{L}\right)^{T}
\end{equation}
with $\left(h\right)_{jk}=J_{j}(\delta_{j,k+1}+\delta_{j+1,k})+V_{j}\delta_{jk}$, $\left(M_{1}\right)_{jk}=\delta_{j1}\delta_{k1}\gamma_{1}$ and
$\left(M_{L}\right)_{jk}=\delta_{jL}\delta_{kL}\gamma_{L}$.
By numerical diagonalization of the damping matrix $X$ for systems
with different $L$, we can explore the size scaling relation of the
Liouvillian gap for the quasiperiodic or disorder chain with boundary
dissipations. 
Before studying the concrete models, we shall use perturbation theory
to derive an analytical expression of the Liouvillian gap under the
weak dissipation limit, which is very helpful for understanding the
scaling relation of Liouvillian gap.

By using Jordan-Wigner transformation to replace
fermion creation and annihilation operators with spin operators, $c_{i}^{\dagger}=P_{i}\sigma_{i}^{+}$, $c_{i}=P_{i}\sigma_{i}^{-}$,
$P_{i}=\prod_{k=1}^{i-1}\sigma_{k}^{z}$, and introducing the Choi-Jamiolkwski
isomorphism \cite{Choi,Jamiolkowski,Tyson,Vidal} which turns the matrix
into a vector:
\[
\rho=\sum_{mn}\rho_{mn}|m\rangle\langle n|\ \rightarrow\ |\rho\rangle=\sum_{mn}\rho_{mn}|m\rangle\otimes|n\rangle,
\]
the Lindblad equation can then be rewritten into the vectorized form
\begin{equation}
\frac{d|\rho(t)\rangle}{dt}=\mathbb{L}|\rho(t)\rangle=(\mathbb{L}_{0}+\mathbb{L}_{1})|\rho(t)\rangle,
\end{equation}
where explicit forms of $\mathbb{L}_{0}$ and $\mathbb{L}_{1}$ are given in the appendix A.

By virtue of the parity operator $Q=\prod_{k=1}^{L}\sigma_{k}^{z}\tau_{k}^{z}$, which satisfies
$[Q,\mathbb{L}]=0$ and has eigenvalues of $\pm1$, we can define
the projection operators $\mathcal{Q}_{\pm}=(1\pm Q)/2$ such
that $\mathbb{L}=\mathbb{L}_{+}\bigoplus\mathbb{L}_{-}=(\mathcal{Q}_{+}\mathbb{L}\mathcal{Q}_{+})\bigoplus(\mathcal{Q}_{-}\mathbb{L}\mathcal{Q}_{-})$.
Since the parity operator only appears in $\mathbb{L}_{1}$, we have $\mathcal{Q}_{+}\mathbb{L}_{0}\mathcal{Q}_{+} =\mathcal{Q}_{-}\mathbb{L}_{0}\mathcal{Q}_{-}= \mathbb{L}_{0}$.
It can be proved that in the specific model we studied,  the Liouvillian gap is not affected by the choice of parity  when only considering perturbation to first-order correction,
so we only need to consider $\mathbb{L}_{+}=\mathbb{L}_{0+}+\mathbb{L}_{1+}$
with
\begin{equation}
\begin{array}{cl}
\mathbb{L}_{0+}=&  \mathbb{L}_{0} =-i\left(\widetilde{H}\otimes\mathbb{I}-\mathbb{I}\otimes\widetilde{H}^{T}\right),\\
\mathbb{L}_{1+}=& \mathcal{Q}_{+}\mathbb{L}_{1}\mathcal{Q}_{+} =\sum_{\mu}\left[2\widetilde{L_{\mu}}\otimes\widetilde{L_{\mu}}^{*}-(\widetilde{L_{\mu}}^{\dagger}\widetilde{L_{\mu}})\otimes\mathbb{I}\right.\\
 & \left.-\mathbb{I}\otimes(\widetilde{L_{\mu}}^{\dagger}\widetilde{L_{\mu}})^{T}\right],\ \ \mu=1,L
\end{array}
\end{equation}
where $\widetilde{H},\ \widetilde{L_{1}},\ \widetilde{L_{L}}$
differ from $H,\ L_{1},\ L_{L}$ only by replacing fermion operators
$c_{i},\ c_{i}^{\dagger}$ with spin operators $\sigma_{i}^{-},\ \sigma_{i}^{+}$
(see Appendix A for details).

Taking $\mathbb{L}_{1+}$
as a perturbation to $\mathbb{L}_{0+}$ and considering
only the first-order perturbation, we assume that the eigenvalues
$\eta_{r,s}^{(0)}$ without perturbation are $d(r,s)$-fold degenerate,
and the corresponding eigenvectors are denoted as set $\{|\Psi_{r,s}\rangle\}$,
where $|\Psi_{r,s}\rangle:=|\psi_{r}\rangle\otimes|\psi_{s}\rangle^{*}$
is the right eigenvector of $\mathbb{L}_{0+}$ with
both $|\psi_{r}\rangle$ and $|\psi_{s}\rangle$ being the eigenvectors
of $\widetilde{H}$. It can be known that the
first-order perturbation to eigenvalues of Liouvillian
superoperator  $\mathbb{L}_{+}$, denoted by $\eta_{r,s}^{(1)}$,
are the eigenvalues of matrix $W$ with matrix elements $W_{k,k^{\prime}}=\langle\Psi_{k}|\mathbb{L}_{1+}|\Psi_{k^{\prime}}\rangle:=\langle\Psi_{r,s}|\mathbb{L}_{1+}|\Psi_{r^{\prime},s^{\prime}}\rangle$,
where $|\Psi_{k^{\prime}}\rangle\equiv|\Psi_{r^{\prime},s^{\prime}}\rangle$
and $|\Psi_{k}\rangle\equiv|\Psi_{r,s}\rangle$ have the same zero
order eigenvalue $\eta_{r,s}^{(0)}$.

Considering $[\widetilde{H},N]=0$, where $N=\sum_{j=1}^{L}\sigma_{i}^{+}\sigma_{i}^{-}$,
we can order the degenerate eigenstates $|\Psi_{r,s}\rangle$ with
the same eigenvalue $\eta_{r,s}^{0}$ from the
smallest to largest in order of $N_{r,s}\equiv\langle\psi_{r}|N|\psi_{r}\rangle+\langle\psi_{s}|N|\psi_{s}\rangle$.
Simple analysis shows that the first term of $\mathbb{L}_{1+}$
has no effect on the eigenvalues of $W$ and thus does not contribute to $\eta_{r,s}^{(1)}$.
Then we obtain the Liouvillian spectrum
\begin{equation}
\eta=i(E_{r}-E_{s})-\sum_{\mu}\gamma_{\mu}(n_{\mu}^{r}+n_{\mu}^{s})
\end{equation}
under the first order approximation and the Liouvillian gap
\begin{equation}
\Delta_{g}=\underset{\eta}{\min}^{\prime}\{\Re(-\eta)\}=2\underset{r}{\min}^{\prime}\{\sum_{\mu}\gamma_{\mu}n_{\mu}^{r}\}, \label{LG1}
\end{equation}
in which both $E_{r}$ and $E_{s}$ being the eigenvalues of the Hamiltonian
and $n_{\mu}^{r}\equiv\langle\psi_{r}|\sigma_{\mu}^{+}\sigma_{\mu}^{-}|\psi_{r}\rangle$,
$\underset{r}{\min}^{\prime}\{x_{r}\}\equiv\underset{r}{\min}\{x_{r}|x_{r}\neq0\}$.
In our model, $\mu=1,L$, it can be seen that the
Liouvillian gap corresponds to the minimum of nonzero sum of $2(\gamma_{1}n_{1}^{r}+ \gamma_{L}n_{L}^{r})$, where $n_{1}^{r}$ ($n_{L}^{r}$) represents the left (right) boundary density of the $r$-th eigenstate of the underlying Hamiltonian $H$. For the case $\gamma_{1}=\gamma_{L}=\gamma$, we have
\begin{equation}
\Delta_{g}= 2\gamma \underset{r}{\min}^{\prime}(n_{1}^{r}+ n_{L}^{r}), \label{LG2}
\end{equation}
which indicates that the Liouvillian gap is proportional to the minimum of boundary densities of eigenstates of the underlying Hamiltonian.

Now we apply Eq.(\ref{LG2}) to give a theoretical interpretation for the different scaling relations of Liouvillian gap in localized and extended phases.  For simplicity, we shall focus on the case of $\gamma_{1}=\gamma_{L}=\gamma$ in the following discussions and calculations.  Eq.(\ref{LG2}) does not rely on the details of underlying Hamiltonian, and the Liouvillian gap is only relevant to the boundary densities of eigenstates of $H$. For the non-interacting Hamiltonian described by Eq.(\ref{eq:H_fermi-1}), solving Liouvillian gap only needs to consider
the single particle space of the Hamiltonian. When
the system is in a localized phase, the modulus of a localized wavefunction can be approximately described by $|\psi_r(j)|\propto e^{-|j-r_{0}|/\xi_r}$, where $r_{0}$ is the index of the localization center and $\xi_r$ is the localization length.
Then the corresponding density distribution is given by $n_{j}^{r}\propto e^{-2\kappa_r \left|j-r_{0}\right|}$, where $\kappa_r=1/\xi_r$ is the Lyapunov exponent of the localized eigenstate. For the quasiperiodic system described by the extended AAH model (see Eq.(\ref{eq:HGAAH})), all eigenstates have the same localization length and Lyapunov exponent, and thus we can denote the state-independent Lyapunov exponent as $\kappa$ (given by Eq.(\ref{LE-AAH}) for the extended AAH model). The different localized eigenstate with the same localization length can be characterized by different localization center $r_{0}$, i.e., $n_{j}^{r}\propto e^{-2\kappa \left|j-r_{0}\right|}$. Then we can estimate the Liouvillian gap by using Eq.(\ref{LG2}), which gives rise to
\begin{equation}
\Delta_{g}\propto2\gamma \underset{r_{0}}{\min}^{\prime}\{ e^{-2\kappa(r_{0}-1)}+ e^{-2\kappa(L-r_{0})}\}\propto\gamma e^{-\kappa L}. \label{eq:local_gap}
\end{equation}
In general, the Lyapunov exponent of a localized eigenstate of quasiperiodic and disordered systems is state-dependent, e.g., the Lyapunov exponent of a localized eigenatate of the quasiperiodic model (\ref{GPD}) is given by Eq.(\ref{eq:Lya_expHop_dual}), which is energy dependent. The Lyapunov exponent $\kappa(E)$ takes its maximum in the top of energy band, and thus applying Eq.(\ref{LG2}) we can estimate
\begin{equation}
\Delta_{g}\propto\gamma e^{-\kappa(E_\text{top}) L}, \label{eq:local_gap2}
\end{equation}
where $E_\text{top}$ represents the eigenvalue of the localized eigenstate on the top of energy band.

Now we apply Eq.(\ref{LG2}) to give a theoretical interpretation for the scaling relation of Liouvillian gap $\Delta_{g} \propto L^{-3}$ in the extended phase. For simplicity, we consider an extreme case of Hamiltonian (\ref{eq:H_fermi-1}) with $J_{i}=1$ and $V_{i}=0$,
then we have $n_{\mu}^{r}=\frac{2}{L+1}\sin^{2}\left(k_{r}\mu\right)$, where
$k_{r}=\frac{r\pi}{L+1}$. By using Eq.(\ref{LG2}), it follows
\begin{equation}
\begin{array}{cl}
\Delta_{g} & =2\gamma(n_{1}^{1}+n_{L}^{1})=\frac{8\gamma}{L+1}\sin^{2}\left(\frac{\pi}{L+1}\right)\\
 & \approx8\gamma\pi^{2}L^{-3}\propto\gamma L^{-3} ,
\end{array}  \label{scalingL3}
\end{equation}
which is consistent with results in references \cite{Prosen2008NJP,Znidaric2015PRE}.

\subsection{Liouvillian gap in boundary-dissipated quasiperiodic and disorder systems}


Our perturbative derivation of Liouvillian gap does not depend on
the details of Hamiltonian. Eq.(\ref{LG2}) suggests that the Liouvillian gap is closely related to the minimum of boundary densities of eigenstates of the underlying Hamiltonian.  As long as $H$ supports localized eigenstates, similar argument holds true by following the procedure of
deriving Eq.(\ref{eq:local_gap2}), and thus
we expect the exponential scaling relation of Liouvillian gap is quite
universal. To get an intuitive understanding, next we numerically study the scaling relation of Liouvillian gap in various boundary-dissipated quasiperiodic and disorder systems with equal boundary dissipation strengthes $\gamma_1=\gamma_L=\gamma$.
\begin{figure*}[t]
\begin{centering}
\includegraphics[scale=0.25]{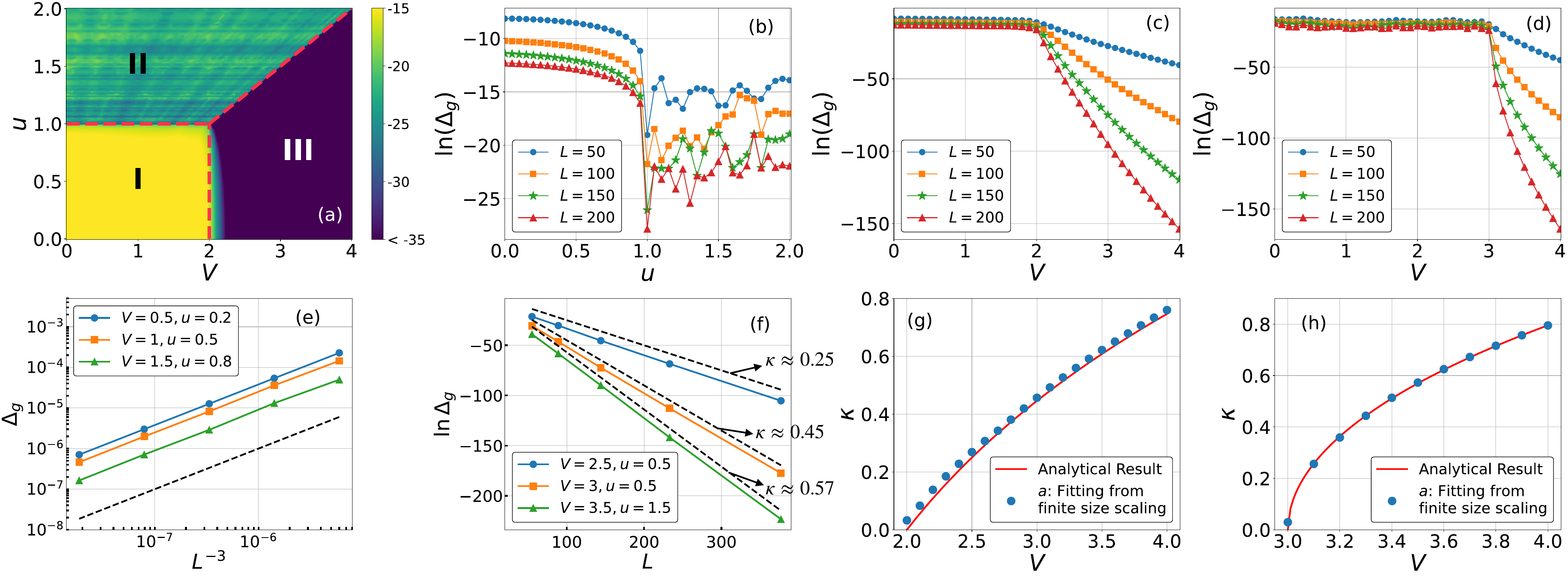}
\par\end{centering}
\caption{\label{fig: Lgap_Vu} (a) $\ln\Delta_{g}$ with respect to $V$ and
$u$ for $L=200$ and $\gamma=1$. The dashed lines denote the phase boundaries of
the underlying phase diagram of the extended AAH model. (b) $\ln\Delta_{g}$
versus $u$ for various size of lattices with $V=0.5$. $\ln\Delta_{g}$
versus $V$ for various size of lattices with (c) $u=0.5$ and (d)
$u=1.5$. Finite size scaling of Liouvillian gap in (e) the extended
phase and (f) the localized phase, where the black dashed lines guide
the value of $\Delta_{g}=L^{-3}$ and $\Delta_{g}=e^{-\kappa L}$,
respectively. Here $L=55,89,144,233,377$ are chosen as the Fibonacci
numbers. Comparing the numerical fitting data $a$ obtained from the
finite size scaling with the analytical result of Lyapunov exponent
for (g) $u=0.5$ and (h) $u=1.5$. The data of (g) and (h) are the
same as (c) and (d) in localized phase, respectively.}
\end{figure*}

To be concrete, we first consider the quasiperiodic system with $H$
described by the extended AAH model \cite{Hatsugai,Takada,HanJH}:
\begin{align}
H= & J\sum_{j=1}^{L-1}\left\{ 1+u\cos\left[2\pi(j+\frac{1}{2})\alpha\right]\right\} \left(c_{j}^{\dagger}c_{j+1}+\text{H.c.}\right)\nonumber \\
 & +V\sum_{j=1}^{L}\cos\left(2\pi j\alpha\right)c_{j}^{\dagger}c_{j},\label{eq:HGAAH}
\end{align}
where $\alpha=(\sqrt{5}-1)/2$, the hopping strength $J$ defines the energy scale
and is set to 1, $c_{j}^{\dagger}(c_{j})$ is the fermion
creation (annihilation) operator, $u$ represents the modulation amplitude
for the off-diagonal hopping, and $V$ is the strength of the on-site
quasiperiodic potential. In the absence of boundary dissipations,
the phase diagram of AAH model is shown in the Fig. \ref{fig: Lgap_Vu}(a)
with the regions I, II and III corresponding to extended, critical,
and localized phases, respectively \cite{Hatsugai,Takada,HanJH}.
The phase boundaries can be obtained with finite-size scaling analyses for the wavefunction properties and
level statistics \cite{Hatsugai,Takada,HanJH}.
For the extended AAH model (\ref{eq:HGAAH}), we note that the Lyapunov exponent can be analytically
expressed as \cite{Jitomirskaya,HanJH}
\begin{equation}
\kappa=\begin{cases}
\max\left\{ \ln\left|\frac{\left|V\right|+\sqrt{V^{2}-4u^{2}}}{2u}\right|,0\right\} , & \left|u\right|\geqslant1\\
\max\left\{ \ln\left|\frac{\left|V\right|+\sqrt{V^{2}-4u^{2}}}{2\left(1+\sqrt{1-u^{2}}\right)}\right|,0\right\} . & \left|u\right|<1
\end{cases}\label{LE-AAH}
\end{equation}
By using the above analytical result, the phase boundaries between localized phase and extended (critical) phase can be analytically determined.

Without loss of generality, we fix $\gamma=1$ and calculate the Liouvillian
gap for various parameters $u$ and $V$. The value of $\ln(\Delta_{g})$
is displayed in the underlying phase diagram in Fig.\ref{fig: Lgap_Vu}(a),
which indicates the Liouvillian gap exhibiting different features
in different phase regions. As shown in Fig. \ref{fig: Lgap_Vu}(b)-(d),
$\ln(\Delta_{g})$ also displays an abrupt change in the phase boundaries
of the underlying phase diagram.
By analyzing the size scaling of $\Delta_{g}$ as shown in Fig.\ref{fig: Lgap_Vu}(e),
we demonstrate that the Liouvillian gap in the extended region fulfills
\begin{equation}
\Delta_{g}(L)\propto L^{-3},
\end{equation}
which is consistent with Eq.(\ref{scalingL3}).
In the critical region, the Liouvillian gap approximately fulfills
the algebraic form
\[
\Delta_{g}(L)\propto L^{-\eta},
\]
where $\eta>3$ is a non-universal exponent sensitive to parameters of $u$ and $V$.
The sensitivity to parameter $u$ can be also witnessed by the oscillation
behavior in Fig.\ref{fig: Lgap_Vu}(b). For the localized phase, the
finite size scaling of $\Delta_{g}$ in Fig.\ref{fig: Lgap_Vu}(f)
shows the Liouvillian gap taking the exponential form:
\begin{equation}
\Delta_{g}(L)\propto e^{-aL},
\end{equation}
where $a$ is a parameter-dependent constant. Our numerical results
unveil that $a$ is identical to the Lyapunov exponent
of the localized phase with $\kappa$ given by Eq.(\ref{LE-AAH}), which is obviously independent of eigenvalues of localized states.
In Fig. \ref{fig: Lgap_Vu}(g) and (h), we plot the Lyapunov exponent versus $V$ according
to Eq. (\ref{LE-AAH}) by taking $u=0.5$ and $1.5$, respectively,
in comparison with the numerical fitting data $a$ obtained from the
finite size scaling, which indicates clearly $a\approx\kappa$ in
the whole underlying localized region.

\begin{figure}
\begin{centering}
\includegraphics[scale=0.23]{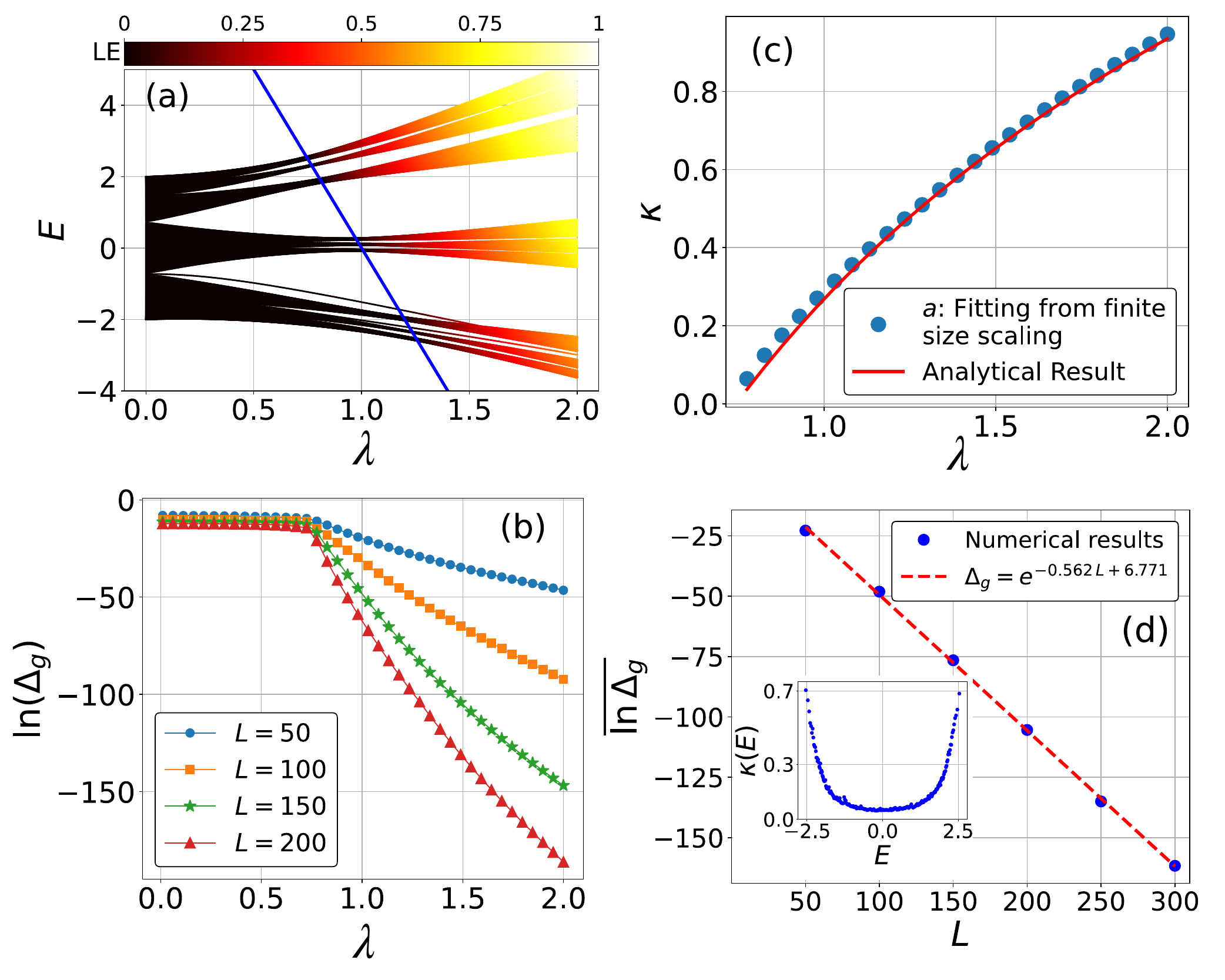}
\par\end{centering}
\caption{\label{fig:expHop_dual} (a) Energy spectrum of Eq.(\ref{GPD}) with
respect to $\lambda$ for $L=200$ with the color representing the
value of Lyapunov exponent of the eigenstate with the corresponding
eigenvalue. The blue solid line represents the exact mobility edge.
(b) $\ln\Delta_{g}$ versus $\lambda$ for various size of lattices
with $b=0.2$; (c) Comparing the numerical fitting data $a$ obtained
from the finite size scaling with the analytical result of Lyapunov
exponent. (d) Finite size scaling of $\ln\Delta_{g}$ for 1D Anderson
model by averaging 100 samples. The insert in (d) shows the Lyapunov
exponent of 1D Anderson model for $L=200$ by averaging 1000 samples.}
\end{figure}

To scrutinize the scaling relation for more complex quasiperiodic systems, next we
consider a quasiperiodic system with a mobility edge described  by the following Hamiltonian \cite{Ganeshan}:
\begin{equation}
H=J\sum_{j=1}^{L-1}\left(c_{j}^{\dagger}c_{j+1}+\text{H.c.}\right)+2\lambda\sum_{j=1}^{L}\frac{\cos\left(2\pi\alpha j\right)}{1-b\cos\left(2\pi\alpha j\right)}c_{j}^{\dagger}c_{j},\label{GPD}
\end{equation}
where $\alpha=(\sqrt{5}-1)/2$ and $b\in(-1,1)$, the hopping strength $J$ defines the energy scale and is set to 1. While Eq. (\ref{GPD})
reduces to the AAH model for $b=0$, the model with $b\neq0$ exhibits
an exact mobility edge following the expression $E=2\,\text{sgn}(\lambda)(1-|\lambda|)/b$.
The Lyapunov exponent for the localized state can be obtained from $\kappa(E)=\max\left\{ \kappa_{c}(E),0\right\} $
with the analytical expression of $\kappa_{c}(E)$ given by \cite{LiuYX2,XiaX}
\begin{equation}
\kappa_{c}(E)=\ln\left|\frac{\left|bE+2\lambda\right|+\sqrt{\left(bE+2\lambda\right)^{2}-4b^{2}}}{2\left(1+\sqrt{1-b^{2}}\right)}\right|,\label{eq:Lya_expHop_dual}
\end{equation}
where $E$ denotes the eigenvalue of Eq. (\ref{GPD}). In Fig. \ref{fig:expHop_dual}(a),
we show the energy spectrum with respect to $\lambda$ of Eq. (\ref{GPD})
with $b=0.2$ and the value of $\kappa(E)$ is denoted by the color.
The mobility edge can be determined by $\kappa_{c}(E)=0$, as illustrated
by the blue solid line in Fig.\ref{fig:expHop_dual}(a), which separates
the extended states from the localized states above it. It can be
seen that the non-zero value of the Lyapunov exponent would appear in spectrum as
$\lambda$ increases across the mobility edge.

By fixing the boundary dissipation strength $\gamma=1$, we display
the Liouvillian gap with respect to $\lambda$ in Fig. \ref{fig:expHop_dual}(b)
for different system sizes. When $\lambda$ exceeds a critical value,
corresponding to the emergence of mobility edge, the size scaling
relation of Liouvillian gap has an obvious change. The finite size
analysis demonstrates that the Liouvillian gap fulfills an exponential
form $\Delta_{g}\propto e^{-aL}$. The exponent $a$ with respect
to $\lambda$ extracted from the exponential fitting of the data is
shown in the Fig. \ref{fig:expHop_dual}(c), which is found to agree
well with $\kappa(E_{\text{top}})$, where $E_{\text{top}}$ denotes
the eigenvalue in the top of the energy band with the corresponding
Lyapunov exponent taking the largest value. It turns out that the size scaling of
Liouvillian gap for this quasiperiodic model can be well described
by $\Delta_{g}\propto e^{-\kappa(E_{\text{top}})L}$, consistent with Eq.(\ref{eq:local_gap2}) as predicted by our theoretical analysis.

Finally, we study the boundary-dissipated 1D Anderson model \cite{Schulz}
with $H$ described by
\begin{equation}
H=J\sum_{j=1}^{L-1}\left(c_{j}^{\dagger}c_{j+1}+\text{H.c.}\right)+\sum_{j=1}^{L}V_{j}c_{j}^{\dagger}c_{j},
\end{equation}
where the on-site random potential $V_{j}$ uniformly distributes
among $[-V,V]$, the hopping strength $J$ defines the energy scale
and is set to 1. For the 1D Anderson model, the state is always localized
for arbitrarily weak disorder strength $V$. By taking $\gamma=1$
and $V=1$, we calculate the Liouvillian gap numerically and find
it also fulfills exponential size scaling relation $\Delta_{g}\propto e^{-aL}$
with $a\approx0.562$, as shown in Fig. \ref{fig:expHop_dual}(d).
As no analytical expression for the Lyapunov exponent of the Anderson
model is available, we can numerically calculate the Lyapunov exponent by using $\kappa\left(E\right)=\ln\left(\max\left(\theta_{i}^{+},\theta_{i}^{-}\right)\right)$,
where $\theta_{i}^{\pm}$ represents eigenvalues of the matrix $\mathbf{\Theta}=\left(T_{L}^{\dag}T_{L}\right)^{1/(2L)}$
and
\[
T_{L}\left(E,\theta\right)=\prod_{j=1}^{L}T^{j}=\prod_{j=1}^{L}\left(\begin{array}{cc}
E-V_{j} & -1\\
1 & 0
\end{array}\right)
\]
is the transfer matrix \cite{LiuYX}. The numerical value of Lyapunov exponent versus
$E$ for $V=1$ is displayed in the inset of Fig. \ref{fig:expHop_dual}(d).
The numerical result indicates that the Lyapunov exponent for the Anderson model
takes its maximum on the band edges. Since the center of localized
wave function randomly distributes on the lattice site, we take an
average over 10 states close to the band edges, which gives a mean
value of Lyapunov exponent $\bar{\kappa}\approx0.589\pm0.066$. It can be seen that
$\bar{\kappa}$ matches well with $a\approx0.562$, i.e., the decaying
exponent can be described by the mean value of Lyapunov exponent close to band edges
of the 1D Anderson model.
\begin{figure}
\begin{centering}
\includegraphics[scale=0.27]{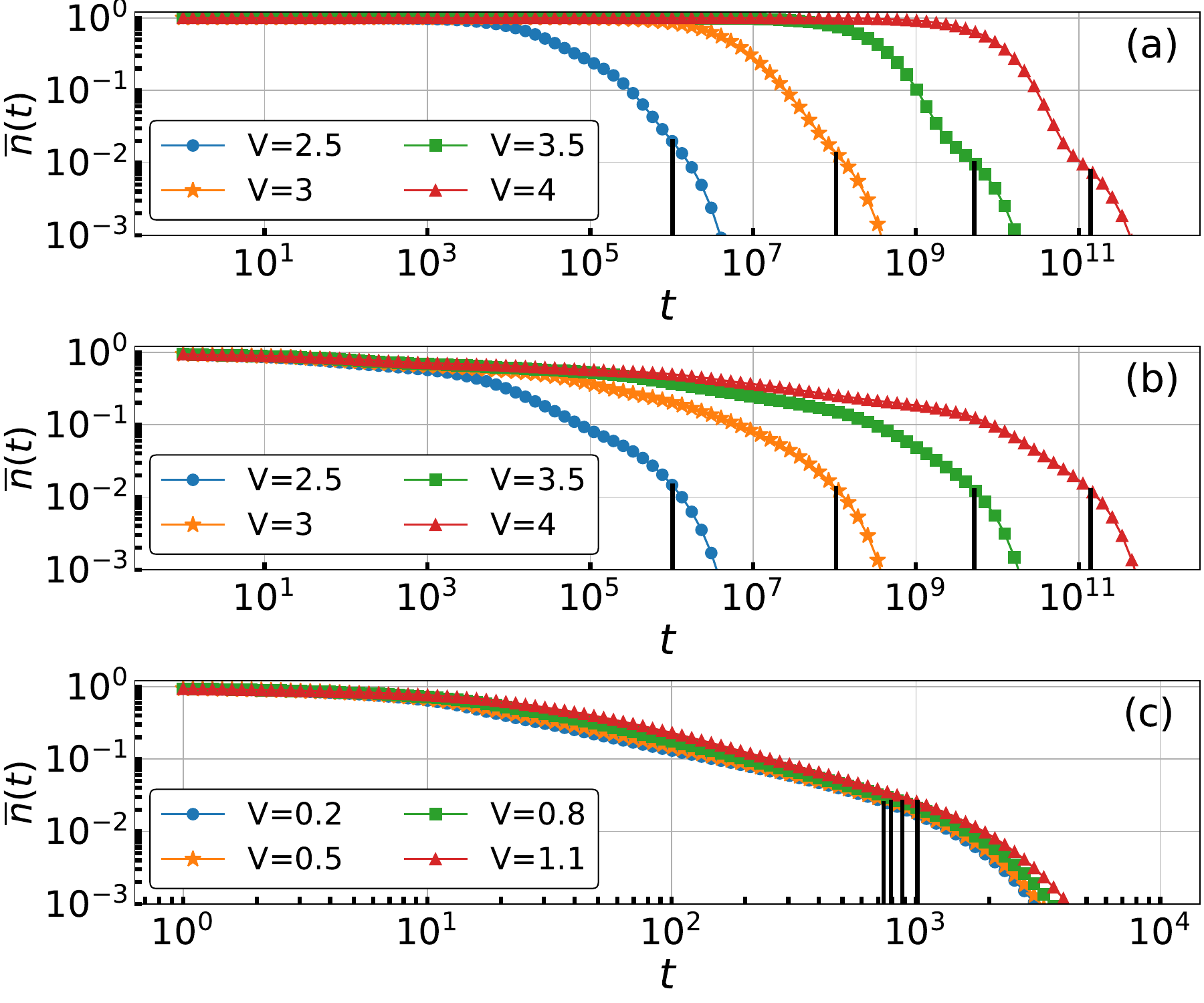}
\par\end{centering}
\caption{\label{fig:Nc_relaxation} The average occupation number $\overline{n}(t)$
in the localized region of boundary-dissipated generalized AAH model
for the initial state chosen as (a) the state localized at the center
site 16; (b) the fully occupied state. (c) $\overline{n}(t)$ in the
extended region with the fully occupied initial state. The black lines
guide values of the inverse of Liouvillian gaps corresponding to different
$V$. Here we have taken $L=30$, $u=0.2$ and $\gamma=1$.}
\end{figure}

\subsection{Relaxation dynamics}

To see clearly how the relaxation timescale related to the Liouvillian gap, we study the dynamical evolution  of the average occupation number for the extended AAH model with boundary
dissipation.  The average occupation number is defined
as $\overline{n}(t)=\sum_{j=1}^{L}\langle n_{j}(t)\rangle/[\sum_{j=1}^{L}\langle n_{j}(t=0)\rangle]$,
where $\langle n_{j}(t)\rangle=\text{Tr}[\rho(t)c_{j}^{\dagger}c_{j}]$.
We demonstrate $\overline{n}(t)$ versus $t$ for the system of $L=30$,
$u=0.2$, $\gamma=1$ and various $V$ with the initial state chosen
as the state localized at the center site $16$ in Fig. \ref{fig:Nc_relaxation}(a)
and a fully occupied state in Fig. \ref{fig:Nc_relaxation}(b), respectively.
For the open system with pure loss dissipation, the nonequilibrium
steady state is the empty state with $\overline{n}(t\rightarrow\infty)=0$.
Since the late-stage dynamics of the system near a steady state is
governed by eigenmodes of Liouvillian whose eigenvalues are close
to zero, the relaxation times can be estimated by the inverse of Liouvillian
gaps, which are labeled by the black lines in the Fig. \ref{fig:Nc_relaxation}
for guidance. It can be observed that the inverse of Liouvillian gap
gives a reasonable timescale for estimating the time of asymptotic convergence to the steady
state. With the increase in $V$, the relaxation time in the localized
phase increases quickly in terms of $\tau\propto e^{\kappa L}$, which
can be approximately represented as $\tau\propto|V|^{L}$ and is much
longer than the relaxation time in the extended state as shown in
Fig. \ref{fig:Nc_relaxation}(c).

\begin{figure}[h]
\begin{centering}
\includegraphics[scale=0.22]{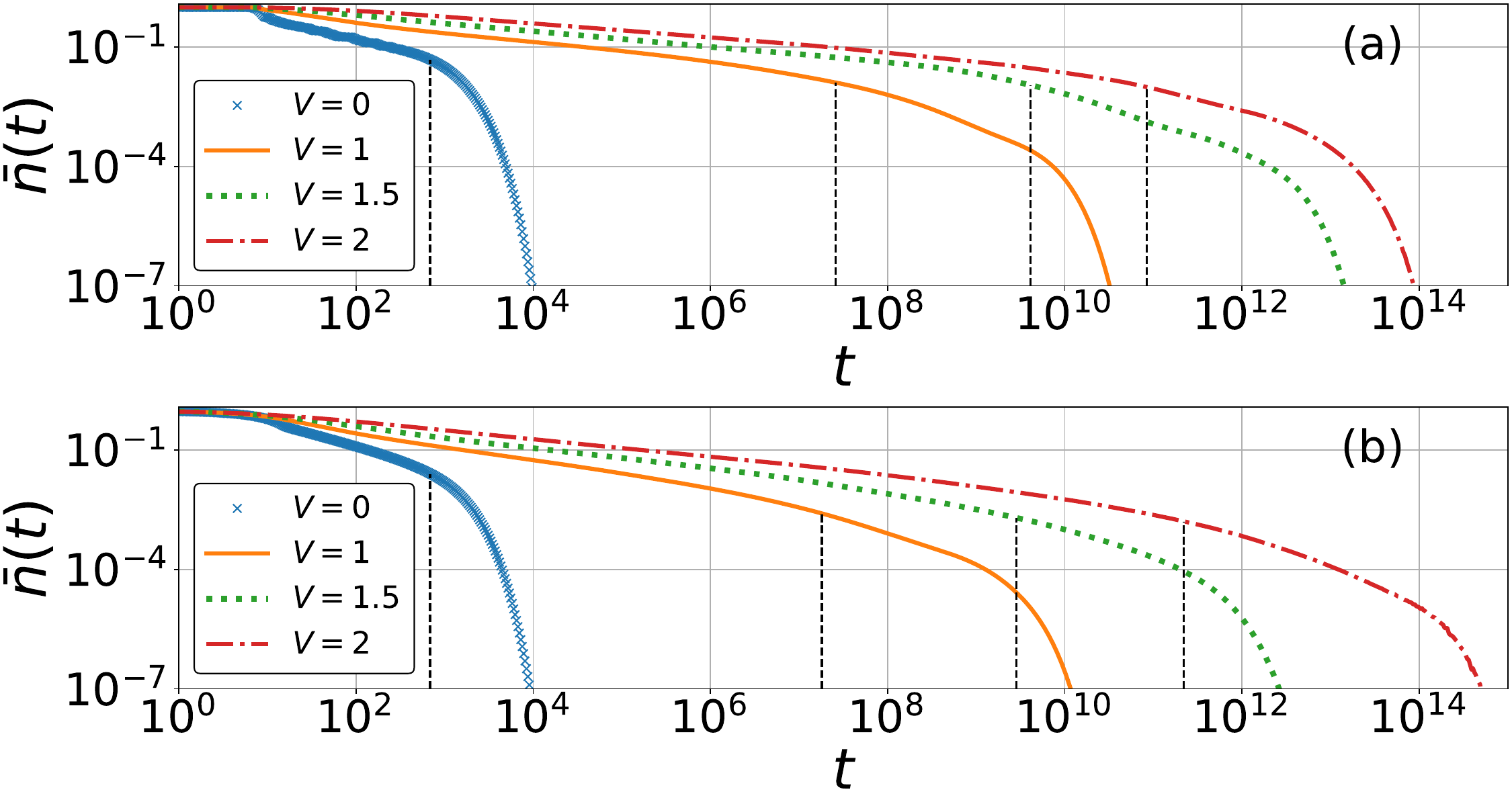}
\par\end{centering}
\caption{\label{fig: nt_anderson} The average occupation number $\overline{n}(t)$
of 1D Anderson model for the initial state chosen as (a) the state
localized at the center site 16; (b) the fully occupied state. The
black lines guide values of the inverse of Liouvillian gaps corresponding
to different $V$. Here we take $L=30$ and averaged 1000 samples
for $V>0$.}
\end{figure}

Next we show the evolution of $\overline{n}(t)$ for the boundary-dissipated 1D Anderson
model with $L=30$, $\gamma=1$ and various $V$. The initial state in Fig. \ref{fig: nt_anderson}(a)
is chosen as the state localized at the center site 16, and in Fig. \ref{fig: nt_anderson}(b) is the fully occupied
state. For guidance, we also mark the values of the inverse of Liouvillian
gaps by the black dashed lines in the figures. The dynamical evolution displays similar behaviors as in the localized phase of the quasiperiodic system. In can be found that the
relaxation time increases quickly as the strength of random potential $V$ increases. Since the states in the 1D Anderson model are always localized, the relaxation time increases exponentially with the increase of system size for any nonzero disorder strength $V$.

\section{Summary and outlook}

In summary, we study the size scaling relation of Liouvillian
gap of boundary-dissipated 1D quasiperiodic and disorder systems  both analytically and numerically.
In the framework of perturbation theory, we give an analytical derivation of the Liouvillian gap by taking the boundary-dissipation terms as a perturbation. Our analytical result unveils that the Liouvillian gap is proportional to the minimum of boundary densities of eigenstates of the underlying Hamiltonian, and thus gives a theoretical explanation why the Liouvillian gap fulfills different size scaling relations when the underlying system is in the extended, critical and localized phase. When the underlying Hamiltonian has localized eigenstates,  the Liouvillian gap displays an exponential size scaling
with the decay exponent determined by the largest Lyapunov exponent of the localized eigenstates.
The exponential size scaling relation was numerically verified in various
quasiperiodic and disorder systems. By studying the dynamical evolution of average occupation number, we show that the inverse of Liouvillian
gap gives a reasonable timescale for estimating the relaxation time.

The quasiperiodic optical lattices have provided an ideal platform
for studying the localization transition in one dimension \cite{Bloch2018,Roati},
and schemes for engineering quasiperiodic optical lattices in open
quantum systems are proposed through purely dissipative processes
\cite{YiW,LiuYG}. Manipulation of laser-induced dissipations \cite{LuoL}
at the boundaries allows us to study the relaxation dynamics of the
quasiperiodic lattices. As the localization length in quasiperiodic
optical lattice can be tuned by engineering the strength of incommensurate
potential, we expect that the relation between the relaxation time
and the localization length of boundary-dissipated quasiperiodic lattice
could be unveiled in the experiment. By considering the interaction effect, it is interesting to study the stability of the many-body localized phase
subjected to boundary dissipation both theoretically \cite{Sels} and experimentally.

\begin{acknowledgments}
We thank Y. X. Liu, C. G. Liang and C. Yang for helpful discussions.
The work is supported by the NSFC under Grants No.12174436 
and No.T2121001 and the Strategic Priority Research Program of Chinese
Academy of Sciences under Grant No. XDB33000000.
\end{acknowledgments}

\appendix

\section{First-order degenerate perturbation of Liouvillian gap}

In this appendix, we give details of the perturbative calculation
of Liouvillian gap.

\subsection{Matrix representation of Liouvillian superoperators}

We consider a dissipative quantum system governed by the Lindblad
equation with the Hamiltonian given by Eq.(\ref{eq:H_fermi-1}) and
the boundary dissipation operators described by the form of Eq.(\ref{eq:LindO}).
Applying the Jordan-Wigner transformation to replace fermion operators
with spin operators, $c_{j}^{\dagger}=P_{j}\sigma_{j}^{+},\ \ c_{j}=P_{j}\sigma_{j}^{-},\ \ P_{j}=\prod_{l=1}^{j-1}\sigma_{l}^{z}$,
we get

\begin{equation}
H^{(spin)}=\sum_{j=1}^{L-1}J_{j}(\sigma_{j}^{+}\sigma_{j+1}^{-}+\sigma_{j+1}^{+}\sigma_{j}^{-})+\sum_{j=1}^{L}V_{j}\sigma_{j}^{+}\sigma_{j}^{-}
\end{equation}

\textcolor{magenta}{{} }
\begin{equation}
L_{1}^{(spin)}=\sqrt{\gamma_{1}}\sigma_{1}^{-}\ \ \ \ ,\ \ \ \ L_{L}^{(spin)}=\sqrt{\gamma_{L}}P_{L}\sigma_{L}^{-}
\end{equation}

In order to give the matrix representation of Liouvillian superoperator,
we introduce the Choi-Jamiolkwski isomorphism that turns the matrix
into a vector: $\rho=\sum_{mn}\rho_{mn}|m\rangle\langle n|\ \ \rightarrow\ \ |\rho\rangle=\sum_{mn}\rho_{mn}|m\rangle\otimes|n\rangle$,
the Lindblad equation can then be rewritten into the vectorized form
$\frac{d|\rho(t)\rangle}{dt}=\mathbb{L}|\rho(t)\rangle=(\mathbb{L}_{0}+\mathbb{L}_{1})|\rho(t)\rangle$
with

\begin{align}
\begin{array}{cl}
\mathbb{L}_{0} & =-i\left(H^{(spin)}\otimes\mathbb{I}-\mathbb{I}\otimes H^{(spin)T}\right)\\
& =-i\left[\sum_{j=1}^{L-1}J_{j}(\sigma_{j}^{+}\sigma_{j+1}^{-}+\sigma_{j+1}^{+}\sigma_{j}^{-}-\tau_{j}^{+}\tau_{j+1}^{-}-\tau_{j+1}^{+}\tau_{j}^{-})\right.\\
& \left.+\sum_{j=1}^{L}V_{j}(\sigma_{j}^{+}\sigma_{j}^{-}-\tau_{j}^{+}\tau_{j}^{-})\right]
\end{array}
\end{align}

\begin{align}
\begin{array}{cl}
\mathbb{L}_{1} & =\sum_{\mu}\left[2L_{\mu}^{(spin)}\otimes L_{\mu}^{(spin)*}-(L_{\mu}^{(spin)\dagger}L_{\mu}^{(spin)})\otimes\mathbb{I}\right.\\
& \left.-\mathbb{I}\otimes(L_{\mu}^{(spin)\dagger}L_{\mu}^{(spin)})^{T}\right]\\
& =2\sigma_{1}^{-}\tau_{1}^{-}+2Q\sigma_{L}^{-}\tau_{L}^{-}-\sum_{\mu=1,L}(\sigma_{\mu}^{+}\sigma_{\mu}^{-}+\tau_{\mu}^{+}\tau_{\mu}^{-})
\end{array}
\end{align}
where $\sigma_{j}^{\alpha},\ \ \tau_{j}^{\alpha}(\alpha=+,-,z)$ are
the Pauli matrices, $Q=\prod_{j=1}^{L}\sigma_{j}^{z}\tau_{j}^{z}$
is the parity operator which satisfies $[Q,\mathbb{L}]=0$. Since the
operator $Q$ has two eigenvalues 1 and -1, we can define the projection
operators $\mathcal{Q}_{+}$, $\mathcal{Q}_{-}$, and divide the Liouville
superoperator space into two parts, thus we have $\mathbb{L}=\mathbb{L}_{+}\bigoplus\mathbb{L}_{-}=(\mathcal{Q}_{+}\mathbb{L}\mathcal{Q}_{+})\bigoplus(\mathcal{Q}_{-}\mathbb{L}\mathcal{Q}_{-})$.
We will see later that if we consider only the first order perturbation,
the part $\sum_{\mu}L_{\mu}^{(spin)}\otimes L_{\mu}^{(spin)*}$ that
parity $Q$ can affect does not contribute to the Liouvillian spectrum,
so we only need to consider $\mathbb{L}_{+}$.

We label $\widetilde{H}=\sum_{j=1}^{L-1}J_{j}(\sigma_{j}^{+}\sigma_{j+1}^{-}+\sigma_{j+1}^{+}\sigma_{j}^{-})+\sum_{j=1}^{L}V_{j}\sigma_{j}^{+}\sigma_{j}^{-},\ \ \widetilde{L_{1}}=\sqrt{\gamma_{1}}\sigma_{1}^{-},\ \ \widetilde{L_{L}}=\sqrt{\gamma_{L}}\sigma_{L}^{-}$,
then we have
\begin{equation}
\begin{array}{cl}
\mathbb{L}_{0+}=\mathcal{Q}_{+}\mathbb{L}_{0}\mathcal{Q}_{+} & =-i\left(\widetilde{H}\otimes\mathbb{I}-\mathbb{I}\otimes\widetilde{H}^{T}\right),\\
\mathbb{L}_{1+}=\mathcal{Q}_{+}\mathbb{L}_{1}\mathcal{Q}_{+} & =\underset{\mu=1,L}{\sum}\left[2\widetilde{L_{\mu}}\otimes\widetilde{L_{\mu}}^{*}-(\widetilde{L_{\mu}}^{\dagger}\widetilde{L_{\mu}})\otimes\mathbb{I}\right.\\
& \left.-\mathbb{I}\otimes(\widetilde{L_{\mu}}^{\dagger}\widetilde{L_{\mu}})^{T}\right]
\end{array}
\end{equation}

The difference between $H,\ L_{1},\ L_{L}$ and $\widetilde{H},\ \widetilde{L_{1}},\ \widetilde{{\color{blue}{\normalcolor L}_{{\normalcolor L}}}}$
is just replacing $c_{j},\ c_{j}^{\dagger}$ with $\sigma_{j}^{-},\ \sigma_{j}^{+}$,
we will drop the superscript $''\sim''$ of $\widetilde{H},\ \widetilde{L_{\mu}}$
in the following discussion.

\subsection{Perturbation theory}

We consider the boundary dissipation term as a perturbation. The unperturbed part
of the Liouvillian is a unitary part, $\mathcal{L}_{0+}:=-i\left[H,\rho\right]$,
while the perturbation term is $\mathcal{L}_{1+}:=\sum_{\mu}\left(2L_{\mu}\rho L_{\mu}^{\dagger}-\left\{ L_{\mu}^{\dagger}L_{\mu},\rho\right\} \right)=\gamma\sum_{\mu}\left(2L_{\mu}^{\prime}\rho L_{\mu}^{\prime\dagger}-\left\{ L_{\mu}^{\prime\dagger}L_{\mu}^{\prime},\rho\right\} \right)$
with $L_{\mu}^{\prime}=L_{\mu}/\sqrt{\gamma}$, where $\gamma$ is a small quantity of dissipative strength, which can be taken as the maximum of $\gamma_\mu$. Here the introduction of a perturbation parameter $\gamma$ is for the purpose of the convenience of perturbation calculation. The vectorized form of the Liouville superoperator $\mathbb{L}_{+}=\mathbb{L}_{0+}+\mathbb{L}_{1+}=\mathbb{L}_{0+}+\gamma\mathbb{L^{\prime}}_{1+}$
can be written as
\begin{align}
\mathbb{L}_{0+}= & -i\left(H\otimes\mathbb{I}-\mathbb{I}\otimes H^{T}\right),\\
\mathbb{L^{\prime}}_{1+}= & \sum_{\mu}\left[2L_{\mu}^{\prime}\otimes L_{\mu}^{\prime*}-(L_{\mu}^{\prime\dagger}L_{\mu}^{\prime})\otimes\mathbb{I}-\mathbb{I}\otimes(L_{\mu}^{\prime\dagger}L_{\mu}^{\prime})^{T}\right]
\end{align}

The right eigenvectors of the unperturbed part $\mathbb{L}_{0+}$
can be written as
\begin{equation}
|\Psi_{r,s}\rangle:=|\psi_{r}\rangle\otimes|\psi_{s}\rangle^{*},
\end{equation}
with both $|\psi_{r}\rangle$ and $|\psi_{s}\rangle$ are the eigenvectors
of the Hamiltonian. The right eigenvalues of $|\Psi_{r,s}\rangle$
are $\eta_{r,s}^{(0)}=i\left(E_{r}-E_{s}\right)$, where $E_{r}$
and $E_{s}$ are the eigenvalues of $H$ with respect to the eigenvectors
$|\psi_{r}\rangle$ and $|\psi_{s}\rangle$, respectively. We assume
that the eigenvalue $\eta_{r,s}^{(0)}$ without perturbation is $d(r,s)$-fold
degenerate, and the corresponding eigenvector is denoted as set $\{|\Psi_{r,s}\rangle\}.$
Let $\mathcal{P}_{0}$ be a projection operator onto the space span
of $\{|\Psi_{r,s}\rangle\}$, $\mathcal{P}_{1}=\mathbf{1}-\mathcal{P}_{0}$
to be the projection onto the remaining states. Let $|\Phi_{r,s}\rangle$
denote the right eigenvectors of\textcolor{blue}{{} }$\mathbb{L}_{+}$
with right eigenvalues $\eta_{r,s}$, i.e.,
\begin{equation}
\mathbb{L}_{+}|\Phi_{r,s}\rangle=\eta_{r,s}|\Phi_{r,s}\rangle.
\end{equation}
Then it follows
\begin{align}
0 & =\left(\eta_{r,s}-\mathbb{L}_{0+}-\gamma\mathbb{L^{\prime}}_{1+}\right)|\Phi_{r,s}\rangle\nonumber \\
& =\left(\eta_{r,s}-\eta_{r,s}^{(0)}-\gamma\mathbb{L^{\prime}}_{1+}\right)\mathcal{P}_{0}|\Phi_{r,s}\rangle \nonumber\\
&
 \ \ \ \ \ +\left(\eta_{r,s}-\mathbb{L}_{0+}-\gamma\mathbb{L^{\prime}}_{1+}\right)\mathcal{P}_{1}|\Phi_{r,s}\rangle.\label{eq:P0P1}
\end{align}
We note that $[\mathcal{P}_{0},\mathbb{L}_{0+}]=0,\ \ [\mathcal{P}_{1},\mathbb{L}_{0+}]=0,\ \ \mathcal{P}_{0}^{2}=\mathcal{P}_{0},\ \ \mathcal{P}_{0}\mathcal{P}_{1}=0$.
By applying $\mathcal{P}_{0}$ and $\mathcal{P}_{1}$ on Eq. (\ref{eq:P0P1})
respectively, we can get two equations:
\begin{align}
\left(\eta_{r,s}-\eta_{r,s}^{(0)}-\gamma\mathcal{P}_{0}\mathbb{L^{\prime}}_{1+}\right)\mathcal{P}_{0}|\Phi_{r,s}\rangle-\gamma\mathcal{P}_{0}\mathbb{L}_{1+}^{\prime}\mathcal{P}_{1}|\Phi_{r,s}\rangle & =0,\label{eq:P0L1P0}\\
-\gamma\mathcal{P}_{1}\mathbb{L^{\prime}}_{1+}\mathcal{P}_{0}|\Phi_{r,s}\rangle+\left(\eta_{r,s}-\mathbb{L}_{0+}-\gamma\mathcal{P}_{1}\mathbb{L^{\prime}}_{1+}\right)\mathcal{P}_{1}|\Phi_{r,s}\rangle & =0.\label{eq: P1L1P1}
\end{align}
Eq. (\ref{eq: P1L1P1}) can be rewritten as
\begin{equation}
\mathcal{P}_{1}|\Phi_{r,s}\rangle=\frac{\gamma\mathcal{P}_{1}\mathbb{L^{\prime}}_{1+}\mathcal{P}_{0}}{\eta_{r,s}-\mathbb{L}_{0+}-\gamma\mathcal{P}_{1}\mathbb{L^{\prime}}_{1+}\mathcal{P}_{1}}|\Phi_{r,s}\rangle.
\end{equation}
Substituting it into Eq. (\ref{eq:P0L1P0}), we get \begin{widetext}
	\begin{equation}
	\left(\eta_{r,s}-\eta_{r,s}^{(0)}-\gamma\mathcal{P}_{0}\mathbb{L^{\prime}}_{1+}\mathcal{P}_{0}-\frac{\gamma^{2}\mathcal{P}_{0}\mathbb{L^{\prime}}_{1+}\mathcal{P}_{1}\mathbb{L^{\prime}}_{1+}\mathcal{P}_{0}}{\eta_{r,s}-\mathbb{L}_{0+}-\gamma\mathcal{P}_{1}\mathbb{L}_{1+}^{\prime}\mathcal{P}_{1}}\right)\mathcal{P}_{0}|\Phi_{r,s}\rangle=0.\label{eq:8}
	\end{equation}
\end{widetext} 
For the eigenvalues to the first order of $\gamma$ and eigenvectors to the zero order,
we obtain
\begin{equation}
\left(\eta_{r,s}-\eta_{r,s}^{(0)}-\gamma\mathcal{P}_{0}\mathbb{L^{\prime}}_{1+}\mathcal{P}_{0}\right)\mathcal{P}_{0}|\Phi_{r,s}\rangle=0.\label{eq:9}
\end{equation}
Define $W=\gamma\mathcal{P}_{0}\mathbb{L^{\prime}}_{1+}\mathcal{P}_{0}=\mathcal{P}_{0}\mathbb{L}_{1+}\mathcal{P}_{0}$
and $\eta_{r,s}^{(1)}=\eta_{r,s}-\eta_{r,s}^{(0)}$, then Eq.(\ref{eq:9})
becomes
\begin{equation}
W\left(\mathcal{P}_{0}|\Phi_{r,s}\rangle\right)=\eta_{r,s}^{(1)}\left(\mathcal{P}_{0}|\Phi_{r,s}\rangle\right)
\end{equation}
The first-order Liouvillian spectrum correction $\eta_{r,s}^{(1)}$
is the eigenvalue of the $d(r,s)$-dimensional square matrix $W$
with matrix elements $W_{k,k^{\prime}}=\langle\Psi_{k}|\mathbb{L}_{1+}|\Psi_{k^{\prime}}\rangle:=\langle\Psi_{r,s}|\mathbb{L}_{1+}|\Psi_{r^{\prime},s^{\prime}}\rangle$.

\subsection{The Liouvillian gap}

We assume $[H,N]=0$, where $N=\sum_{j=1}^{L}\sigma_{j}^{+}\sigma_{j}^{-}$
and $L$ is the system size, then the eigenstates of Hamiltonian have
a definite total number of particles. We can label the eigenstates
of the Hamiltonian in terms of energy eigenvalues, total number of
particles, and other expected values of physical quantities: $\left|\psi_{r}\right\rangle =\left|E_{r},N_{r},...\right\rangle ,\ \ r=1,2,...,2^{L}$.

Considering the case with all dissipations taking the form of loss:
$L_{\mu}=\sqrt{\gamma_{\mu}}\sigma_{\mu}^{-}$ , we have
\begin{equation}
(L_{\mu}\otimes L_{\mu}^{*})|\Psi_{r,s}\rangle=L_{\mu}|\psi_{r}\rangle\otimes L_{\mu}^{*}|\psi_{s}\rangle^{*}=\underset{r^{\prime},s^{\prime}}{\sum}g_{r^{\prime},s^{\prime}}|\Psi_{r^{\prime},s^{\prime}}\rangle.\label{eq:L_kron_L}
\end{equation}
The operators $L_{\mu}$ will reduce the particle number of state
$|\psi_{s}\rangle$, and $|\Psi_{r,s}\rangle:=|\psi_{r}\rangle\otimes|\psi_{s}\rangle^{*}$
has a fixed total particle number $N_{r,s}=N_{r}+N_{s}$. Using formula
(\ref{eq:L_kron_L}), we have $N_{r^{\prime},s^{\prime}}<N_{r,s}$.
We can order the degenerate eigenstates $|\Psi_{r,s}\rangle$ with
the same eigenvalue $\eta_{r,s}^{(0)}$ from the smallest to largest
in order of $N_{r,s}$. For convenience, we relabel $|\Psi_{k}\rangle:=|\Psi_{r,s}\rangle$
with the double index $r,s$ replaced by a new index $k$, and $N_{r^{\prime},s^{\prime}}<N_{r,s}$
can be substituted by $k^{\prime}<k$. So only if $k^{\prime}<k$,
we have $\langle\Psi_{k^{\prime}}|(L_{\mu}\otimes L_{\mu}^{*})|\Psi_{k}\rangle\neq0$.

If $\eta_{k}^{(0)}=i(E_{r}-E_{s})=i(E_{r^{\prime}}-E_{s^{\prime}})=\eta_{k^{\prime}}^{(0)}$,
assume that the eigenvalues of Hamiltonian has no degeneracy, then
we have $\delta_{r,r^{\prime}}=\delta_{s,s^{\prime}}=\delta_{k,k^{\prime}}$.
Labeling $n_{\mu}^{r}=\langle\psi_{r}|\sigma_{\mu}^{+}\sigma_{\mu}^{-}|\psi_{r}\rangle$,
then we have
\begin{widetext}
	\begin{equation} \langle\Psi_{k^{\prime}}|[I\otimes(L_{\mu}^{\dagger}L_{\mu})^{T}]|\Psi_{k}\rangle=\langle\psi_{r^{\prime}}|\psi_{r}\rangle\left(\langle\psi_{s^{\prime}}|(L_{\mu}^{\dagger}L_{\mu})^{\dagger}|\psi_{s}\rangle\right)^{*}=\delta_{k,k^{\prime}}\gamma_{\mu}n_{\mu}^{s},
	\end{equation}
	\begin{equation}
	\langle\Psi_{k^{\prime}}|[(L_{\mu}^{\dagger}L_{\mu})\otimes I]|\Psi_{k}\rangle=\langle\psi_{r^{\prime}}|(L_{\mu}^{\dagger}L_{\mu})|\psi_{r}\rangle\left(\langle\psi_{s^{\prime}}|\psi_{s}\rangle\right)^{*}=\delta_{k,k^{\prime}}\gamma_{\mu}n_{\mu}^{r}.
	\end{equation}
\end{widetext}

It turns out that $W$ is an upper triangular matrix with eigenvalues
of $\eta_{r,s}^{(1)}=-\sum_{\mu}\gamma_{\mu}(n_{\mu}^{r}+n_{\mu}^{s})$.
Since the effect of $\sum_{\mu}L_{\mu}\otimes L_{\mu}^{*}$
	appears in the off-diagonal part of $W$, the effect of different
	parity is not reflected in the first-order perturbation correction
	of the Liouvillian spectrum, but in the higher-order perturbation
	correction.

We obtain the first-order modified Liouvillian spectrum
\begin{equation}
\eta=i(E_{r}-E_{s})-\sum_{\mu}\gamma_{\mu}(n_{\mu}^{r}+n_{\mu}^{s})
\end{equation}
and Liouvillian gap
\begin{equation}
\Delta_{g}=\underset{\eta}{\min}^{\prime}\{\Re(-\eta)\}=2\underset{r}{\min}^{\prime}\{\sum_{\mu}\gamma_{\mu}n_{\mu}^{r}\},
\end{equation}
where $\underset{r}{\min}^{\prime}\{x_{r}\}\equiv\underset{r}{\min}\{x_{r}|x_{r}\neq0\}$
means taking the minimum among all nonzero elements of $x_{r}$.

If all dissipations take the form of gain, $L_{\mu}=\sqrt{\gamma_{\mu}}\sigma_{\mu}^{+}$,
following the similar calculation, we have
\begin{equation}
\left\{ \begin{array}{c}
\eta=i(E_{r}-E_{s})+\sum_{\mu}\gamma_{\mu}(n_{\mu}^{r}+n_{\mu}^{s}-2),\\
\Delta_{g}=2\underset{r}{\min^{\prime}}\{\sum_{\mu}\gamma_{\mu}(1-n_{\mu}^{r})\}.
\end{array}\right.
\end{equation}

In the situation that we are considering here, we can see that the
Liouvillian eigenvalue, which determines the Liouvillian gap, is given
by adding perturbation to the zero eigenvalue of $\mathbb{L}_{0}$.

\textbf{Lemma 1:} Given a one-dimensional Hermitian quadratic Hamiltonian
$H$ composed of fermions (or bosons), its single particle eigenvalues
and eigenstates are denoted as $\varepsilon_{j}$ and $\left|\varphi_{j}\right\rangle $,
respectively. We select a sequence $\overrightarrow{\nu}=(\begin{array}{cccc}
\nu_{1}, & \nu_{2}, & \cdots & ,\nu_{L}\end{array})$ with $\nu_{j}\in\{0,1\}$(or $\nu_{j}\in\mathbb{N}$) and label the
multiparticle eigenstate corresponding to the eigenvalue $E_{\overrightarrow{\nu}}\equiv\stackrel[j=1]{L}{\sum}(\nu_{j}\varepsilon_{j})$
of $H$ as $\left|\varphi_{\overrightarrow{\nu}}\right\rangle $,
then we have $\forall m\in\{1,2,...,L\},\ \ \left\langle \varphi_{\overrightarrow{\nu}}\right|c_{m}^{\dagger}c_{m}\left|\varphi_{\overrightarrow{\nu}}\right\rangle =\stackrel[j=1]{L}{\sum}\nu_{j}\left\langle \varphi_{j}\right|c_{m}^{\dagger}c_{m}\left|\varphi_{j}\right\rangle $.

According to the Lemma 1, when the dissipation terms are only loss,
solving Liouvillian gap only need to consider the single particle
space of the Hamiltonian. For the GAA model in the localized phase,
we have $n_{\mu}^{j}\propto e^{-2\kappa\left|\mu-j_{0}\right|}$.
Considering the dissipation $L_{1}=\sqrt{\gamma}c_{1}$ and $L_{L}=\sqrt{\gamma}c_{L}$,
we get
\begin{equation}
\begin{array}{cl}
\Delta_{g} & \propto2\underset{j_{0}}{\min}^{\prime}\{\gamma e^{-2\kappa(j_{0}-1)}+\gamma e^{-2\kappa(L-j_{0})}\}\\
& =\left\{ \begin{array}{c}
4\gamma e^{\kappa}e^{-\kappa L},\ \ \text{when \ensuremath{L} is odd,}\\
2\gamma(1+e^{\kappa})e^{-\kappa L},\ \ \text{when \ensuremath{L} is even,}
\end{array}\right.
\end{array}
\end{equation}
which gives rise to $\Delta_{g}\propto\gamma e^{-\kappa L}$ for any
$L$.

Similar analyses can be carried out for the extended phase. Consider
the limit case of the extended AAH model with $V=u=0$, for which
the expectation value of a local density operator for the $j$-th eigenstate under open boundary condition is
given by $n_{\mu}^{j}=\frac{2}{L+1}\sin^{2}\left(k_{j}\mu\right)$, where
$k_{j}=\frac{j\pi}{L+1}$ with $j=1,\cdots,L$ and $\mu$ is the label
of site. It can be found that the boundary density at $\mu=1$ and $\mu=L$
is minimum for $j=1$ or $L$, i.e.,
\begin{equation}
\Delta_{g}=2\gamma(n_{1}^{1}+n_{L}^{1})=\frac{8\gamma}{L+1}\sin^{2}\left(\frac{\pi}{L+1}\right)\approx8\gamma\pi^{2}L^{-3} .
\end{equation}
The last approximation holds if $L$ is large enough. This derivation gives an explanation why the Liouvillian gap for
the extended state scales in terms of $\Delta_{g}\propto\gamma L^{-3}$.

Now we give the proof of the Lemma 1\textbf{:} We consider that the
Hamiltonian has quadratic fermionic (or bosonic) form:
\begin{equation}
H=\stackrel[l,j=1]{L}{\sum}h_{l,j}{\color{magenta}{\normalcolor c_{l}^{\dagger}c_{j}}},
\end{equation}
where $h$ can be diagonalized with matrix $P$ constructed from a
single particle eigenvector $\left|\varphi_{j}\right\rangle $:
\begin{equation}
h=P\Lambda P^{-1},\ \ P=\left[\begin{array}{cccc}
\left|\varphi_{1}\right\rangle  & \left|\varphi_{2}\right\rangle  & \cdots & \left|\varphi_{L}\right\rangle \end{array}\right].
\end{equation}
The Hermitian property of the Hamiltonian guarantees that $P^{-1}=P^{\dagger}$.
The Hamiltonian can be written as a diagonal form in the new fermion(or
boson) operator $d_{j}$,
\begin{equation}
H=\sum_{j}\varepsilon_{j}d_{j}^{\dagger}d_{j},
\end{equation}
where we denote $\varepsilon_{j}$ as energy eigenvalues which are
the entries of the diagonal matrix $\Lambda$ and ${\color{magenta}{\normalcolor c}}_{m}=\sum_{j}P_{mj}d_{j}$.

In the $d$-fermion (or boson) representation, the many-particle eigenvector
can be written as
\begin{equation}
\begin{array}{cl}
\left|\varphi_{\overrightarrow{\nu}}\right\rangle  & :=\left|\nu_{1},\cdots,\nu_{L}\right\rangle \\
& =\left[\stackrel[j=1]{L}{\Pi}\frac{(d_{j}^{\dagger})^{\nu_{j}}}{\sqrt{\nu_{j}!}}\right]\left|0\right\rangle ,\ \nu_{j}\in\{0,1\}\ (or\ \nu_{j}\in\mathbb{N}),
\end{array}
\end{equation}
with the eigenvalue $E_{\overrightarrow{\nu}}=\stackrel[j=1]{L}{\sum}(\nu_{j}\varepsilon_{j})$
and $\left|0\right\rangle $ is the vacuum state. Then the occupation
number of the many-particle state can be calculated via
\begin{widetext}
	\[
	\begin{array}{cl}
	\left\langle \varphi_{\overrightarrow{\nu}}\right|c_{m}^{\dagger}c_{m}\left|\varphi_{\overrightarrow{\nu}}\right\rangle  & =\stackrel[l,j=1]{L}{\sum}P_{l,m}^{\dagger}P_{m,j}\left\langle \nu_{1},\cdots,\nu_{L}\right|d_{l}^{\dagger}d_{j}\left|\nu_{1},\cdots,\nu_{L}\right\rangle \\
	& =\stackrel[l,j=1]{L}{\sum}P_{l,m}^{\dagger}P_{m,j}\delta_{l,j}\nu_{j}\\
	& =\stackrel[j=1]{L}{\sum}\nu_{j}P_{j,m}^{\dagger}P_{m,j}\\
	& =\stackrel[j=1]{L}{\sum}\nu_{j}\left\langle \varphi_{j}\right|c_{m}^{\dagger}c_{m}\left|\varphi_{j}\right\rangle .
	\end{array}
	\]
\end{widetext}

\end{document}